\documentclass[11pt,onecolumn,draftclsnofoot]{IEEEtran}
\usepackage{amsfonts}
\IEEEoverridecommandlockouts

\ifCLASSINFOpdf
\else
\fi
\usepackage{psfig}
\usepackage{epsf}
\usepackage{graphicx}
\usepackage{bm} 
\usepackage{slashbox}
\usepackage{booktabs}
\usepackage{pifont}
\usepackage{subfigmat} 
\usepackage{times,amsmath,color,amssymb,graphicx,geometry, epsfig,psfrag,subfigure,algorithm}
\usepackage{amssymb,amsmath}
\usepackage{enumerate}
\usepackage{color}

\begin{document}
\title{ Unmanned Aerial Vehicle Swarm-Enabled Edge Computing: Potentials, Promising Technologies, and Challenges}
\author{\IEEEauthorblockN{Wei Wu, \emph{Member, IEEE}, Fuhui Zhou, \emph{Senior Member, IEEE}, Baoyun Wang, \emph{Member, IEEE}, Qihui Wu, \emph{Senior Member, IEEE}, Chao Dong, and Rose Qingyang Hu, \emph{Fellow, IEEE}}
\thanks{This work was supported in part by National Key R${\rm{\& }}$D Program of China under Grant 2020YFB1807602, in part by the National Natural Science Foundation of China under Grant 61901231, Grant 62071223, Grant 62031012, Grant 61931011 and Young Elite Scientist Sponsorship Program by CAST, in part by the China Postdoctoral Science Foundation under Grant 2020M671480 and Postdoctoral Science Foundation of Jiangsu Province under Grant 2020Z295, in part by the open project of the Key Laboratory of Dynamic Cognitive System of Electromagnetic Spectrum Space, Ministry of Industry and Information Technology under Grant KF20202102, in part by the National Key Scientific Instrument and Equipment Development Project under Grant 61827801.}

\thanks{W. Wu is with the College of Communication and Information Engineering, Nanjing University of Posts and Telecommunications,
Nanjing, 210003, China, and with the Key Laboratory of Dynamic Cognitive System of Electromagnetic Spectrum Space, Ministry of Industry and Information Technology, Nanjing University of Aeronautics and Astronautics, Nanjing, 211106, China (e-mail: weiwu@njupt.edu.cn)

F. Zhou, Q. Wu, and C. Dong are with College of Electronic and Information Engineering, Nanjing University of Aeronautics and Astronautics, Nanjing, 211106, China (e-mail: zhoufuhui@ieee.org, wuqihui2014@sina.com, dch999@gmail.com). \textit{(Corresponding author: Fuhui Zhou.)}

B. Wang is with the College of Communication and Information Engineering, Nanjing University of Posts and Telecommunications, Nanjing, 210023, China (e-mail: bywang@njupt.edu.cn).

R. Q. Hu is with the Department of Electrical and Computer Engineering, Utah State University, Logan, UT 84322 USA (e-mail:rosehu@ieee.org).
}

}
\maketitle
\begin{abstract}
Unmanned aerial vehicle (UAV) swarm enabled edge computing  is envisioned to be promising in the sixth generation wireless communication networks due to their wide application sensories and flexible deployment. However, most of the existing works focus on edge computing enabled by a single or a small scale UAVs, which are very different from UAV swarm-enabled edge computing. In order to facilitate the practical applications of UAV swarm-enabled edge computing, the state of the art research is presented in this article. The potential applications, architectures and implementation considerations are illustrated. Moreover, the promising enabling technologies for UAV swarm-enabled edge computing are discussed. Furthermore, we outline challenges and open issues in order to shed light on the future research directions.

\end{abstract}
\begin{IEEEkeywords}
Unmanned aerial vehicle swarm-enabled  edge computing, state of the art, promising technologies, implementation considerations, challenges.
\end{IEEEkeywords}
\IEEEpeerreviewmaketitle
\section{Introduction}



In the past few years, unmanned aerial vehicles (UAVs) have been extensively increasing and widely applied in diverse areas, such as military, agriculture as well as commerce and public services. Moreover, inspired by the swarming behaviors of wild creatures in nature, it is envisioned that more and more UAVs appear in formation or in swarm in practical applications, including the Low-cost Swarm Technology Project initiated by United States in 2015, Jungle Wolf UAVs exhibited at the Farnborough Air Show in 2016, and the successful group flight test of 119 UAVs by China Electronics Technology Group Corporation in 2017 \cite{F. Zhou-WC20}. Furthermore, it is expected that UAVs  will have full autonomous clustering ability by 2025, which will provide the UAV swarm with powerful capability \cite{Y. Zeng-WC19}.

Besides UAV technologies, edge computing is promising since it can significantly improve the computation capacity of the finite-energy and computation-constrained mobile devices. However, traditional ground edge computing networks cannot work in the emergency scenario, such as nature disasters. Recently, the integration of edge computing (MEC) technology into UAV-enabled networks has attracted great attention \cite{F. Zhou-WC20}. A UAV can be considered as a user/relay/MEC-server by properly using computing resources at the edge of network to decrease processing latency and to improve the utilization efficiency, and the performance of UAV-enabled edge computing network can be significantly enhanced. Compared to the conventional one single UAV and small-scale UAVs enabled edge computing networks, the \textit{UAV swarm-enabled edge computing} networks can tackle more complex missions, such as cargo delivery, earth monitoring, precision agriculture and large-scale military deployment, which will gain wide popularity in supporting future human activities. It is envisioned that UAV swarm-enabled edge computing will provide strong support for us to embrace the forthcoming era of ``Internet of Drones (IoD)'' \cite{Y. Zeng-WC19, X. Wang-Access19}. This promising technology has many appealing advantages, presented as follows.



\textit{Significantly Improved Task Execution Capability}: UAV swarm opens up new opportunities for  coordinated multiple points (CoMP) techniques, which enable the computation resources to be shared among a large number of UAVs. By designing a network level resource sharing scheme such that the tasks can be efficiently processed in parallel, UAV swarm-enabled edge computing networks can significantly improve task execution capability \cite{F. Zhou-WC20, X. Wang-Access19}.

\textit{Enhanced Offloading Security}: Due to the flexible aerial CoMP, multiple single-antenna UAVs in a swarm can be combined to form a virtual multiple-antenna system, with which the signal reception quality at the desired receiver sides can be significantly enhanced with the same or even less transmission power \cite{X. Wang-Access19}. As a result, the task offloading physical layer security in the UAV swarm-enabled edge computing networks can be guaranteed.


\textit{High Fault Tolerance}: With on-board sensors and valid authentication mechanism, the UAV swarm system can possess high fault tolerance \cite{Q. Zhang-TVT20}. For example, if several UAVs leave a swarm, the operation can still be maintained with the rest ones forming the reconstructed flying network.

\textit{Helpful UAV-to-UAV Communications}: Last but not least, similar to ground device-to-device (D2D) communications, UAV-to-UAV communications can be widely used in the intra-swarm aerial wireless network for data transfer, relaying, autonomous flying, gathering, and so on \cite{M. Azari-TWC20}. As a result, it will bring benefits in energy efficiency, spectrum efficiency, and extended aerial coverage. Moreover, the burden of backhaul link can be relieved, resulting in lower transmission latency.

The research and development on UAV swarm-enabled edge computing are still in its infancy, and there have been very limited related studies \cite{Q. Zhang-TVT20, J. Zheng-TII20, J. Chen-GLOBECOM19}. The authors in \cite{Q. Zhang-TVT20} studied the response delay minimization issue for a swarm of three-dimensional distributed UAVs through joint optimization of communications and computation resources. To avoid the malicious use of UAV swarm-enabled edge computing system, the authors in \cite{J. Zheng-TII20} investigated the accurate detection and localization of UAV swarms. A service-driven collaborative MEC model was further investigated in \cite{J. Chen-GLOBECOM19} to support computation-intensive and delay-critical services in UAV swarm. To provide a comprehensive understanding and promote the in-depth research, in this paper, we provide an overview on UAV swarm-enabled edge computing. Some key implementation issues are highlighted and the promising technologies are discussed. Furthermore, we shed light on the challenges in UAV swarm-enabled edge computing networks and outline the open research issues.

The rest of this article is organized as follows. The main application scenarios and the basic underpinning architecture paradigms are presented in Section II. The key implementation considerations are elaborated in Section III. Section IV presents the promising technologies to enable UAV swarm edge computing The challenges and open issues for future research are outlined in section V. The article is concluded in Section VI.

\section{State of The Art}
In this section, the state-of-the-art studies on UAV swarm-enabled edge computing are presented. The potential application scenarios and UAV swarm-enabled edge computing architectures are first discussed.
\subsection{Application Scenarios}
\begin{figure}[!t]
\centering
\includegraphics[width=5 in]{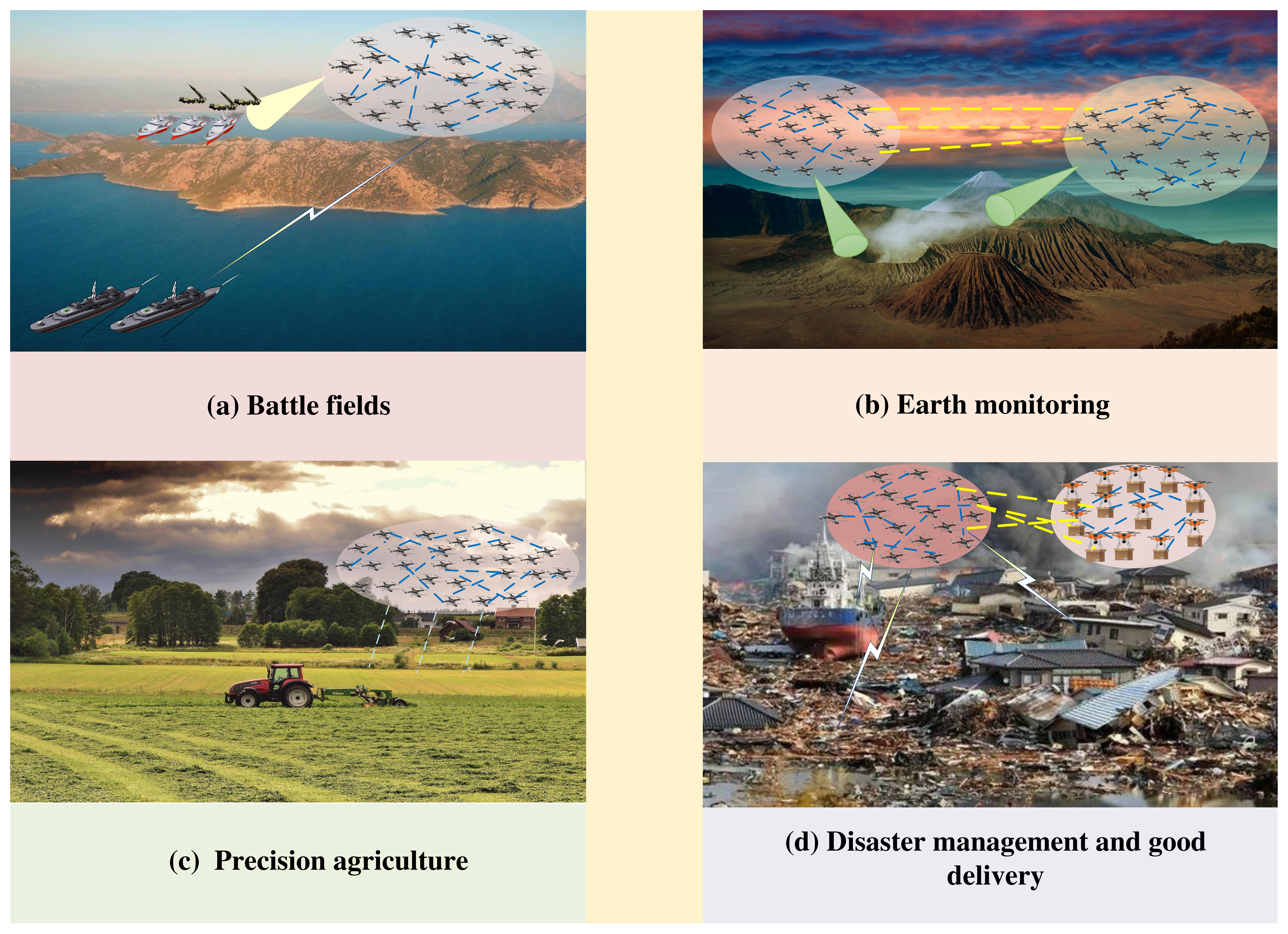}
\caption{Potential application scenarios of UAV swarm-enabled MEC networks.} \label{fig.1}
\end{figure}
\subsubsection{Battle fields} In the battle field, the UAV swarm covering a large area can gather and process massive real-time intelligence, and then transmit them to the nearby weapon stations, greatly enhancing their battlefield situational awareness. For example, UAV swarm can be used to help locate the hidden enemies in street battles and find enemy ships out of sight in naval battles in a real time manner. UAV swarm with aerial computing can be highly instrumental in these scenarios.

\subsubsection{Earth monitoring} The traditional geomatics mechanisms on monitoring the ongoing geophysical processes, such as volcanoes and sea dynamics, and capturing atmospheric temperature, pollutant levels, carbon emissions in the air, etc, are usually costly or time consuming due to the lack of nearby edge computing infrastructures. As an efficient alternative, the deployment of a fleet of drones equipped with on-board sensors and processing units can perform the above tasks timely and cost-effectively by autonomously flying along the predetermined path.

\subsubsection{Precision agriculture} In precision agriculture, on one hand, it is time consuming and prone to risks to perform health monitoring and spraying pesticides to the diseased plants by humans. On the other hand, it is expensive for the farmers to build infrastructures over a wide range of agricultural areas. In this case, UAV swarm is appropriate for enhancing productivity and cost efficiency by performing hyper spectral imagery, data analysis and spraying cooperatively with the on-board relevant devices.

\subsubsection{Disaster management and good delivery} In the disaster situation, it is crucial to carry out rescue in the initial several hours. The lack of reliable communication connection and basic emergency medical supplies will undoubtedly bring adverse impact on lives and properties. Therefore, the situation normally requires the rescue forces to provide immediate assistance. For this case, UAV swarm with aerial computing can play a very important role. On one hand, UAVs can help to evaluate the risks and damages in time by using the built-in computing processing functions, as well as providing connectivity to isolated human-beings. On the other hand, UAV swarm-based faster and cheaper cargo delivery services have attracted many attentions. Amazon Prime Air and Google's Wing projects are the pioneers in this filed using online computing.

\subsection{Recent Advances}

\begin{figure}
  \centering
  \subfigure[UAV equipped with computing units.]{\includegraphics[width=4 in]{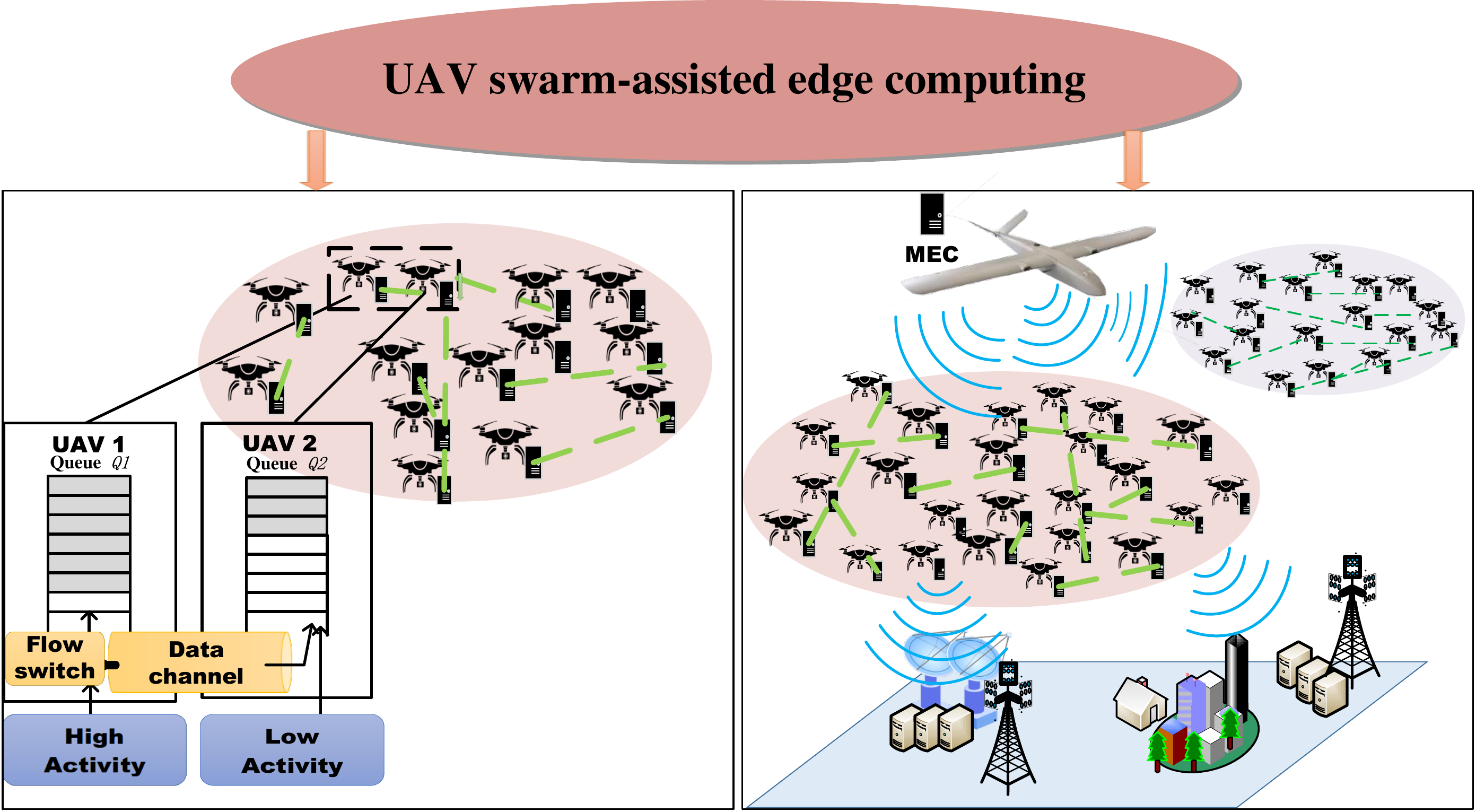}}
  \subfigure[UAV without computing units.]{\includegraphics[width=4 in]{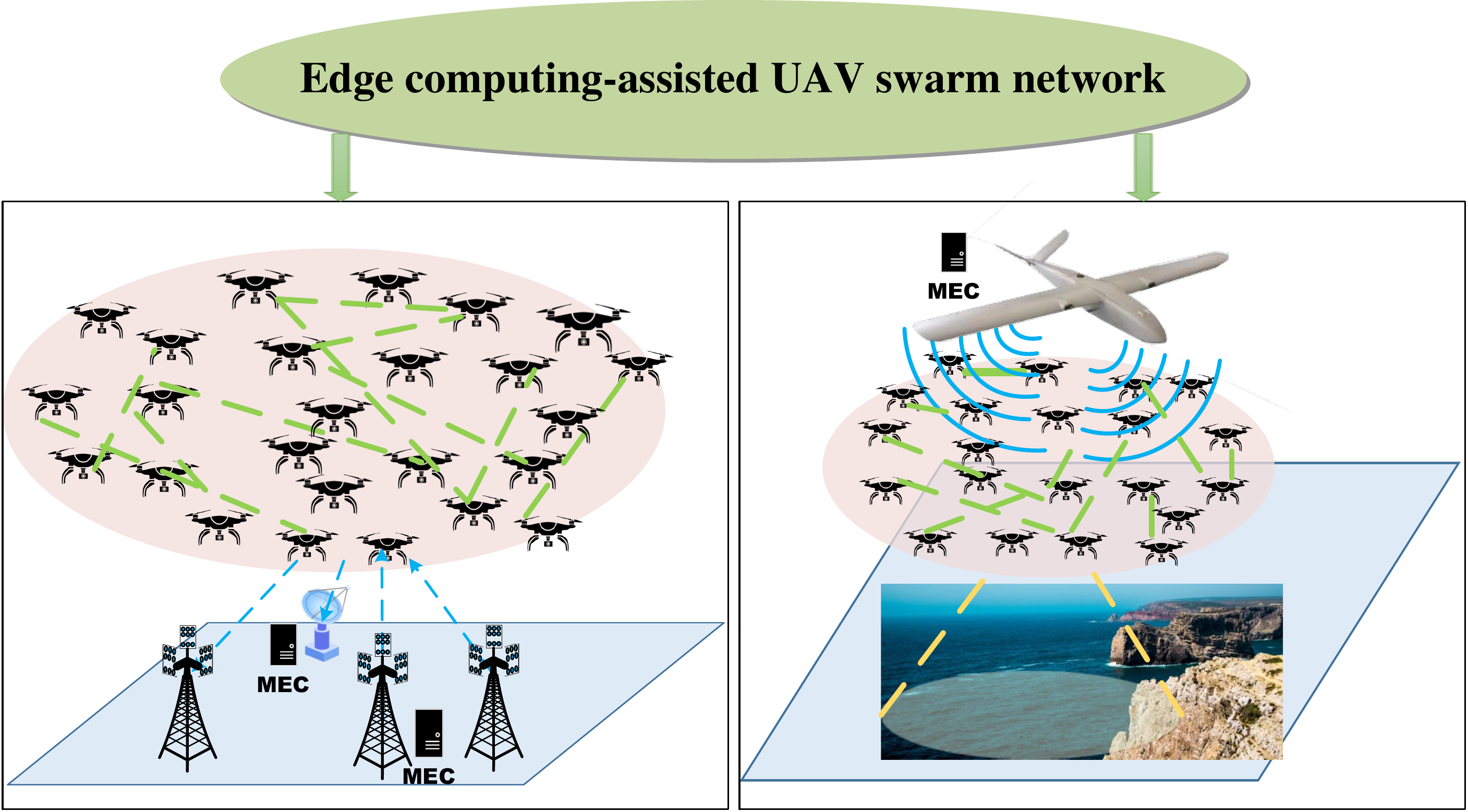}}
  \caption{ Two paradigms of UAV swarm-enabled edge computing.} \label{fig.2}
\end{figure}

Note that among the recent UAV swarm-enabled edge computing applications, UAV swarm acting as the MEC server has attracted extensive research attention \cite{G. Faraci-JSAC20}-\cite{S. Wan-IoT20}. Under this paradigm, dedicated UAVs in the swarm are equipped with computing elements to assist data processing locally or through fine-grained offloading, which we refer to as \textit{UAV swarm-assisted edge computing}, as illustrated in Fig. \ref{fig.2}(a). Besides, in another paradigm, UAVs may always play the role of users and seek for the computing service from nearby server due to their limited sizes and capabilities \cite{Q. Zhang-TVT20}, which we refer as \textit{edge computing-assisted UAV swarm network}, as illustrated in Fig. \ref{fig.2}(b).

In particular, for the first paradigm, to meet the ubiquitous ultra-low latency service requirements of 5G system applications, the authors in \cite{G. Faraci-JSAC20} proposed a Flying Ad-Hoc Network (FANET) consisting of the UAVs equipped with MEC facilities to extend the network slices. Under the proposed framework, the large volumes of data produced by ground devices located in geographic areas that are far away from the structured core network can reach computing services in time. Moreover, each UAV's computing elements can be adaptively turned on and off depending on the activity level of the area being monitored. If one of the UAVs is overload, it can offload jobs to the nearby under-loaded UAVs with some probability to conserve its energy and hence prolong the flight duration of the UAV swarm. However, to achieve a higher accuracy for a larger amount of data, the further analysis besides initial processing on UAVs is necessary \cite{B. Yang-IoT20, S. Wan-IoT20}. Different from the partial and full offloading modes \cite{F. Zhou-WC20}, as the third operation mode in MEC, the data here was calculated and compressed twice. In particular, to identify and track the ground moving target from the captured video frames, the authors in \cite{B. Yang-IoT20} proposed a hierarchical deep learning tasks distribution framework including the lower layers and the higher layers of convolutional neural network (CNN). The lower layers were embedded at each UAV for a preliminary inference of the captured images due to the limited computation resources available in UAV, while the higher layers were deployed at the close-by ground MEC server (GMES) for further processing the intermediate data offloaded from UAV swarm. It achieved a higher inference accuracy by utilizing more powerful computation resources of GMES at the cost of delay. Similarly, the authors in \cite{S. Wan-IoT20} further proposed a three-layer online big data processing network. As the middle layer components, clusters of UAVs were used to collect raw data from the terrestrial distributed sensors in the bottom layer. After the initial steps of data processing, the UAVs then sent the processed results to a center cloud in the top layer for further evaluation. To alleviate the constraints of limited onboard energy on edge computing, dynamic computation and communication resource allocation strategy was employed among UAVs.

For the second paradigm, a swarm of UAVs can be used in the scenarios of environmental monitoring, disaster rescue, fire-fighting, and so on. Note that the applications usually take place in the remote areas or somewhere out of the coverage of terrestrial cellular network. In this case, the authors in \cite{Q. Zhang-TVT20} proposed a two-hop UAV architecture for transferring the important and urgent messages to the control center as quick as possible. During the first hop, the small rotary-wing UAVs in the swarm offload the detected data to the big fixed-wing UAV through wireless uplink. The fixed-wing UAV compresses the received data by onboard MEC server and then forwards the results to the control center via the backhaul link during the second hop. Using a large-size aircraft as aerial computing platform is a promising alternative especially in the areas that are not covered by ground networks.


\section{Implementation Considerations}

By collaborating with each other, UAV swarm-enabled edge computing can open new opportunities for accomplishing more complicated missions. However, it also calls for a paradigm shift in the joint design of computation and communication resource allocation to enable the efficient task execution. The following implementation considerations are presented.

\subsection{Unique Channel Characteristics}

Different from traditional one single UAV-enabled edge computing systems, the UAV swarm-enabled edge computing systems have unique wireless channel characteristics owing to the complex task environment and topology.

In the edge computing systems with one or multiple UAVs, the relatively static line-of-sight (LoS) channel model has been widely used for both the UAV-to-ground and UAV-to-UAV channels. However, it cannot be applied to the UAV swarm scenarios. Note that although LoS path exists due to the high flying altitude of UAV swarm, the scattered paths still exist and cannot be ignored. The scattering caused by ground obstacles may be evaded by UAVs' flexible movements, but that introduced by other UAVs in the swarm is inevitable. Therefore, for the UAV-to-ground channels in the UAV swarm-enabled edge computing systems, both the LoS and non-line-of-sight (NLoS) propagations will be experienced of necessity \cite{Q. Zhang-TVT20, Y. Han-TWC20}. For the UAV-to-UAV channels, the small-scale path fading dominates the channel model, because the information transmitted by one UAV is severely scattered with high probability by many other UAVs in the swarm \cite{M. Azari-TWC20}.

Such unique channel characteristics present new design considerations. On one hand, the LoS and NLoS probability based channel model should be considered for the UAV-to-ground channels, since the LoS probability typically increases with the UAV swarm altitude. On the other hand, the UAV swarm density and UAVs' relative speeds based UAV-to-UAV channel models should be taken into account, since blockage and movement are two of the main factors affecting small-scale fading.

\subsection{Computation Ability vs UAV Swarm Density}

A larger fleet of UAV swarm can lead to a wider area coverage and more diverse mission capabilities, which results in a heavier burden to the edge computing servers. To meet the rigorous service requirement in terms of execution delay, parallel computing can be introduced to support simultaneous multi-task calculation. In particular, by creating many virtual machines (VMs) on the same physical machine (PM), the computing servers' ability can be significantly enhanced. As a result, the queuing delay can be notably reduced and the data congestion is greatly alleviated. Then, a denser UAV swarm is allowed. This can be verified by the result given in \cite{Q. Zhang-TVT20}. Compared to the PM with a single VM, the queueing delay of PM with multiple VMs can be reduced by up to 90 percent for the scenario that a top-UAV with centralized MEC supports the heavy payload of a swarm of distributed bottom-UAVs. However, the number of VMs needs to be carefully controlled. On one hand, it is a big challenge for UAVs to provide sustainable energy to a large number of VMs. On the other hand, the severe I/O interference reduces the individual VM's computation rate and thus compromises the marginal gain from parallel computing. In addition, a denser UAV swarm may result in stronger intra-group interference, which reduces data successful transmission probability and thus brings about higher transmission delay due to the retransmission process.

\subsection{Routing Protocols for UAV Swarm}

\begin{figure}[!t]
\centering
\includegraphics[width=5 in]{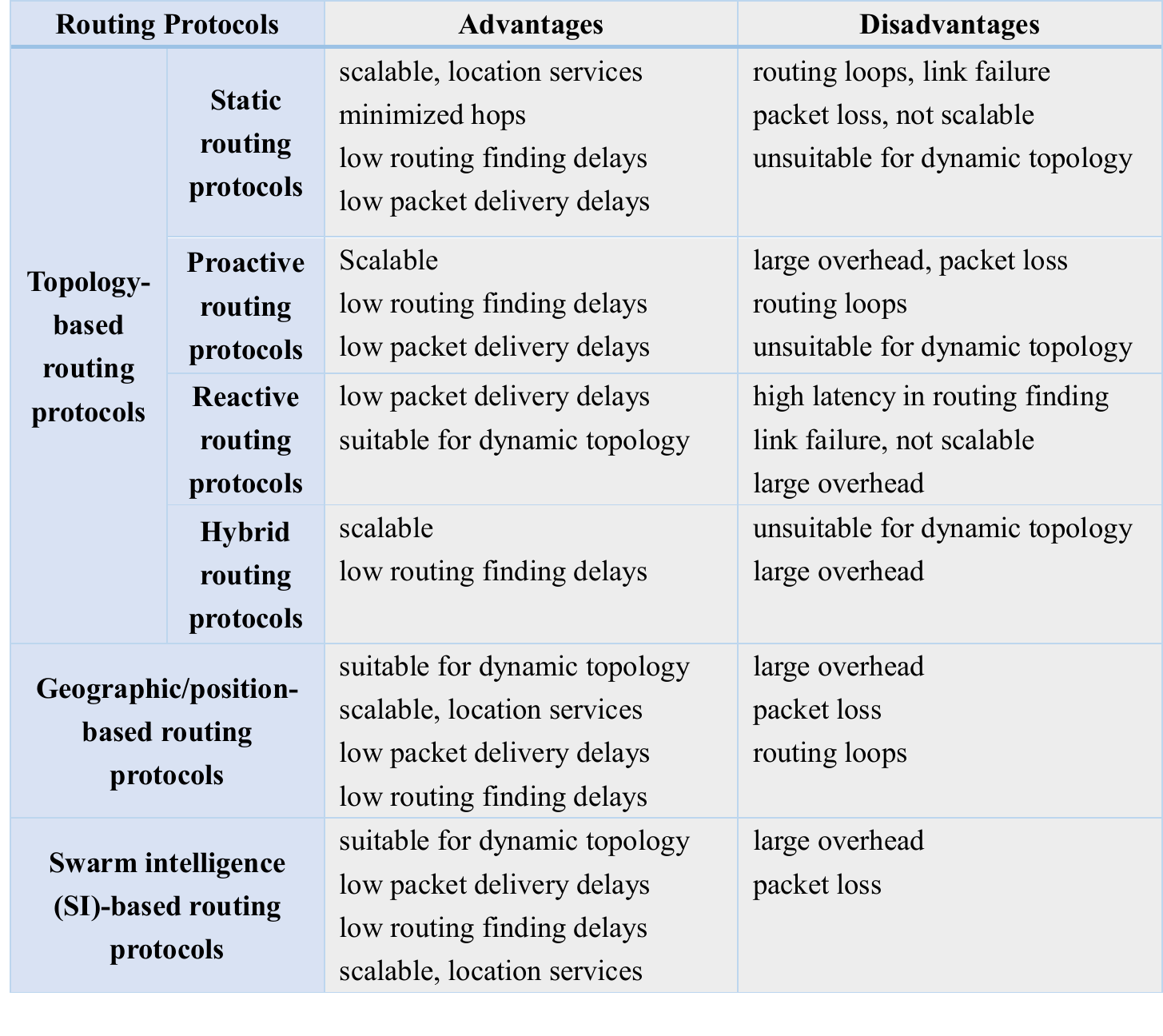}
\caption{The enabling routing protocols for UAV swarm communications.} \label{fig.5}
\end{figure}

Routing protocol plays a vital role in enabling reliable UAV-to-UAV communications for UAV swarm. However, due to the specialty of UAVs, such as fast velocity, intermittent air links, diversified QoS demands and limited energy, traditional ad hoc routing protocols cannot be used directly in intra-swarm or inter-swarm communication. As stated above, UAV formation can be organized as the FANET, which shares many similarities with traditional ad hoc network. Therefore, traditional routing protocols can be utilized as the basis to provide valuable reference for the construction of the routing protocols for UAV swarm.

By modifying the basic routing technologies and combination of several routing technologies, some recently formed protocols are available in three paradigms \cite{X. Chen-AS20}, as illustrated in Fig. \ref{fig.5}. Topology-based routing protocols can be adopted when the UAV swarm-based FANET is in a relatively static state such that the routing table is available. Geographic/position-based routing protocols can be introduced when the UAV swarm is highly dynamic so that the routing table is difficult to obtain. In this paradigm, location services are available to facilitate proper routing. Swarm intelligence (SI)-based routing protocols are inspired by the swarming patterns of bees, birds, ants, etc. By treating each UAV in the swarm as an agent, the routing can be performed with the aid of corresponding algorithm.

\section{Enabling Promising Technologies}

In this section, we discuss several promising technologies to efficiently enable future UAV swarm edge computing networks.

\subsection{Coordinated Multi-Point (CoMP) Computing}

Collaborative computing among multiple UAVs in a swarm or the GMESs is an effective technique to enhance the swarm's computing capability and reduce computation latency by offloading tasks to under-loaded UAVs or GMES. Specifically, for the case that UAVs are equipped with computing units, when some of them are overloaded, they ask help from the closest UAVs that are not stressed at that moment in the swarm. For the case that UAVs have no computing units, they can offload all the  computation tasks to their associated GMESs \cite{F. Zhou-WC20}. A local resource manager is required for coordinating, allocating, and releasing resources on request, which could be the head UAV or the remote system controller with a sufficiently low response latency.

Although multi-UAV cooperation has been studied in conventional UAV-enabled communications, the CoMP computing in UAV swarm still remains uninvestigated, which calls for the joint design of MEC and UAV-enabled wireless network. Moreover, the complicated environment makes it difficult to model the problem accurately and solve it online effectively. Machine learning (ML) is deemed a powerful mechanism in dealing with this problem. 


\subsection{Multiple Access for Task Offloading}

\textbf{Non-Orthogonal Multiple Access (NOMA):} By resorting to successive interference cancellation at the receiver side, NOMA can significantly improve the spectrum efficiency in 5G and future B5G communication systems. Such a high performance gain is achieved under the condition that the channel conditions of users have sufficient diversity. This makes NOMA a very attractive technology for the multi-GMES cooperation scenarios where the computation tasks for each UAV are jointly received and processed by multiple cooperating GMESs, thanks to their different wireless propagation environments due to geographic location of GMESs. For the scenarios that one GMES serves multiple UAVs, NOMA may not be attractive since UAVs in the same swarm normally  have similar channel characteristics with the serving GMES.

\textbf{Space-Division Multiple Access (SDMA):} For the scenarios that one GMES serves multiple UAVs, SDMA is a promising alternative for task offloading. In particular, the GMES can be equipped with multiple phased array antennas to generate different beams concurrently to support the SDMA communication with UAVs. In each phase array antenna penal, multiple radio frequency circuits (RFCs) should be deployed such that each UAV can be assigned a dedicated main lobe beam of GMES. Moreover, for the special cases that many UAVs are located within the same GMES beam due to their embedded inaccurate location awareness techniques, the conventional orthogonal multiple access (OMA) technology, such as the time-division multiple access (TDMA) or the frequency-division multiple access (FDMA), can be utilized as a supplement to further avoid the intra-group interference.


\subsection{Resource Allocation}

Resource allocation is of vital importance in UAV swarm-enabled edge computing networks due to limited available communication and computation resources. A good resource allocation scheme can effectively help UAV swarm to enhance its task execution ability in a cost-effective way.

Due to the coupling between spectrum resources and computing resources, as well as the binary UAV-task association, the resource allocation problem is generally formulated as a mixed-integer nonlinear programming (MINLP)  problem. Some existing approaches like branch and bound method as well as exhaustive search can solve the MINLP problem effectively and obtain the global optimal resource allocation scheme, but with high complexity. Hence, it is better to design an offline approach in a centralized optimization manner, which calls for prior knowledge of the network, such as locations of UAV swarm and tasks, computation task size, number of offloading requests, and computation unit capacity.

As computation complexity of the traditional optimization algorithm increases exponentially with the swarm size and is computationally prohibitive for large-scale UAVs, heuristic local searching (e.g., coordinate decent and convex relaxation) is a good choice in terms of reducing computation complexity. However, it has the drawbacks that the performance cannot be guaranteed and the scheme  is unsuitable for real-time processing in the rapidly changing environment. Therefore, a non-real-time suboptimal resource allocation scheme can be obtained with the heuristic local search methods. 

For large-scale UAVs scenario, as an alternative, learning based resource allocation is a promising technology to manage multi-dimensional resources (e.g., spectrum, computing, and caching resources) in UAV swarm-enabled edge computing with verified low computation complexity. Based on it, the best offloading decision, UAV-task association and trajectory planning can be obtained in a low-delay and high energy efficient manner. The actor-critic architecture based deep reinforcement learning (DRL) can be taken as an example \cite{L. Wang-TCCN20}. The deep deterministic policy gradient (DDPG) technique, consisting of deep neural network (DNN) and deep Q network (DQN), can help UAV solve continuous-valued control problems in the high-dimensional action spaces. It enables the online trajectory design and offloading decisions of UAV swarm. The training of DNN and DQN is performed at the cloud center while the execution is decentralized at each UAV of so-called agent.

\subsection{Spectrum Management}

The basic communication requirements of UAV can be classified into two categories: control and non-payload communication (CNPC) and payload communication. CNPC refers to the two-way communications between UAV and control station to guarantee the safe, reliable and effective flight operation, while payload communication refers to mission-related information transmission.

For CNPC, the dedicatedly protected aviation spectrum has to be assigned due to its vital importance. According to the report by International Telecommunications Union (ITU), a maximum of $34$ MHz terrestrial spectrum and $56$ MHz satellite spectrum is needed to support CNPC for future a large number of UAVs \cite{Y. Zeng-WC19, ITU-09}. As a result, the C-band spectrum at $5030$-$5091$ MHz has been assigned to UAV CNPC at the WRC-12.

For payload communication, there are static spectrum management (SSM) and dynamic spectrum management (DSM) that can be applied to different scenarios. For the scenarios that UAV swarm stays in a certain mission area for a long time, the spectrum usage time is long and constant. In this case, the SSM can be adopted to allocate the unlicensed spectrum to UAV swarm in a centralized manner. For the scenarios that UAV swarm flies fast through different mission areas, the spectrum usage is temporary in a certain area. In this case, the DSM can be employed such that UAVs can share the common spectrum pool with ground users. Cognitive radio is a promising technology in DSM, with which the UAV swarm is able to access the available spectrum opportunistically or in a shared form as long as the interference to primary network is properly controlled.


\subsection{Pilot Decontamination}

\begin{figure}[!t]
\centering
\includegraphics[width=5 in]{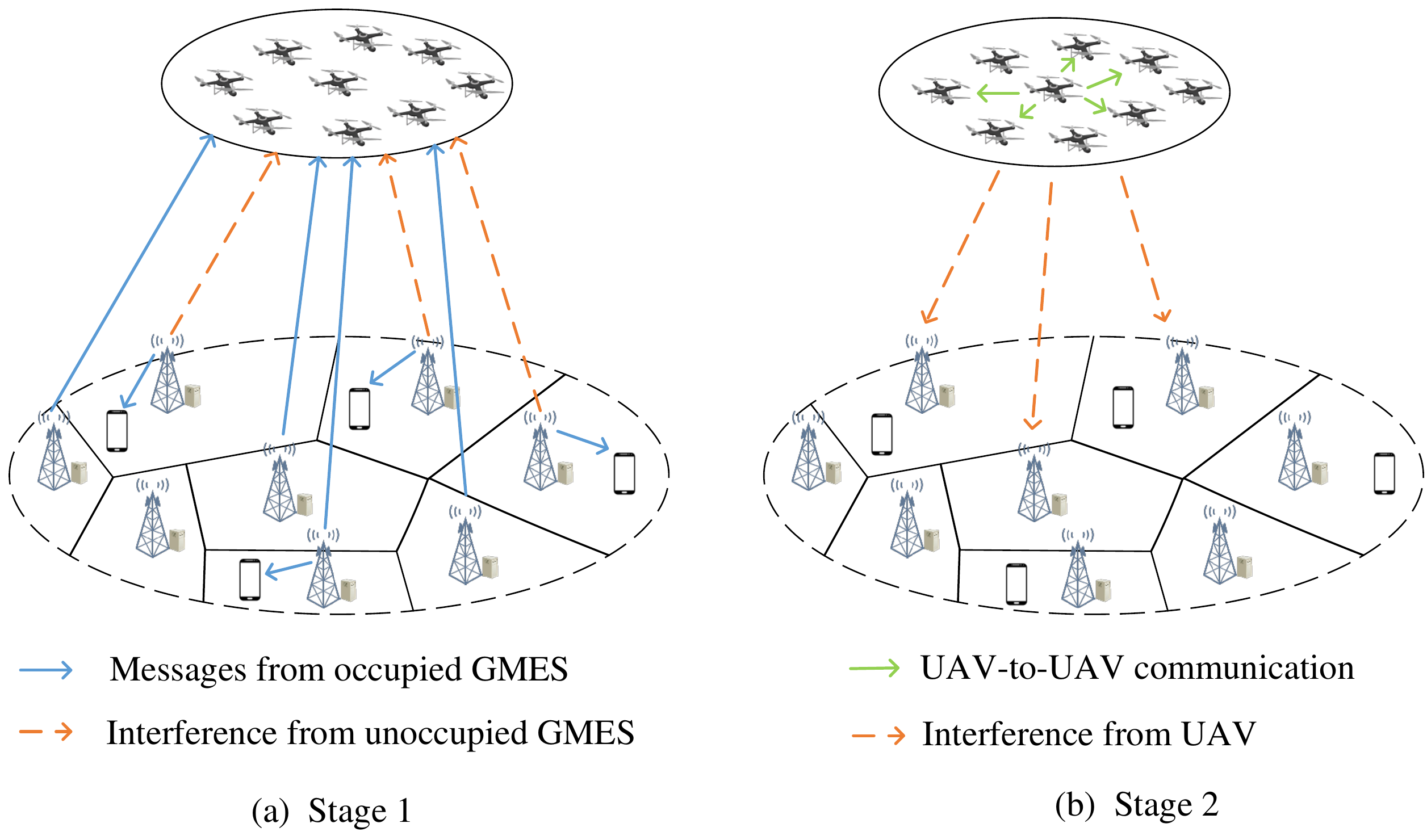}
\caption{D2D communication among UAVs with two-stage transmission for reducing pilot contamination.} \label{fig.4}
\end{figure}

Different from the terrestrial wireless communications, the pilot contamination of UAV-enabled wireless communications is much more complicated due to UAV's higher working altitude and wider coverage area. With the channels mainly dominated by LoS propagation, one UAV can cause severe pilot contamination to ground users and surrounding UAVs that reuse the same pilot \cite{Y. Huang-arxiv20}. Such a problem becomes more challenging to solve in the UAV swarm systems with a large group of coordinated UAVs.

For the pilot contamination on ground users, reserving a dedicated pool of pilots solely possessed by UAVs is an effective approach to decontaminate it when the number of UAVs in a swarm is not too high. When the number of UAVs is large, the spatial interference cancellation techniques can be implemented at each GMES independently.

Compared to the pilot contamination on ground users, the pilot contamination among intra-group UAVs is more challenging due to their similar channel characteristics with GMESs or aerial computing platforms. In this case, D2D communication based two-stage transmission is a promising technology to substantially reduce the pilot overhead and thus the pilot contamination, as illustrated in Fig. \ref{fig.4}. In stage one, one swarm head UAV is chosen ahead for channel estimation with the GMES by sending a pilot in the uplink. Then, based on the estimated single CSI, the occupied GMES forms beams toward the head UAV for sending message to it. In stage two, using D2D communication, the head UAV broadcasts the received message to the following UAVs in the swarm, with which they can retrieve their intended information. In this way, the pilot contamination among UAVs and the corresponding pilot overhead are dramatically decreased, since it is degenerated to the case as one single UAV-enabled edge computing. Even so, due to the required minimum safe distance among UAVs, the practical irrelevance between head UAV-ground channel and following UAV-ground channel should be taken into account. Moreover, multiple head UAVs can also be appointed to ease the burden of single head UAV and reduce transmission delay with the careful design of mutual interference.

\subsection{Energy Consumption Management}

Due to the limited battery capacity, energy consumption is one of the important factors that impacts the performance of UAV. Most of the existing research works only impose the roughly constraints on flight range, flight time and total power budget when scheduling the UAVs. It may result in the insufficient on-board UAV battery power and cannot complete the scheduled task. Therefore, it is of crucial importance to accurately estimate the impact of the communication/computation payload, the flying speed and other factors on the battery energy consumption. Several efficient energy consumption management approaches can be summarized as follows.

1). Building the linear models of payload and energy consumption. Based on it, the battery weight and size can be optimized beforehand to improve the energy conversion efficiency.

2). Introducing the concept of battery consumption rate (BCR). BCR is a key index that affects the battery duration of UAV. Most of the schemes cannot work due to low batteries without considering BCR.

3). Detailing the flying states of UAV. By dividing the flying states of UAV into take-off, cruise, landing and hovering, we can develop more detailed energy allocation schemes for different states and flight rates to increase energy efficiency.

\section{Challenges and Open Issues}

In this section, we outline some critical challenges and open issues that will be encountered in future research of UAV swarm-enabled edge computing.

\subsection{Interference Alleviation}

\begin{figure}[!t]
\centering
\includegraphics[width=5 in]{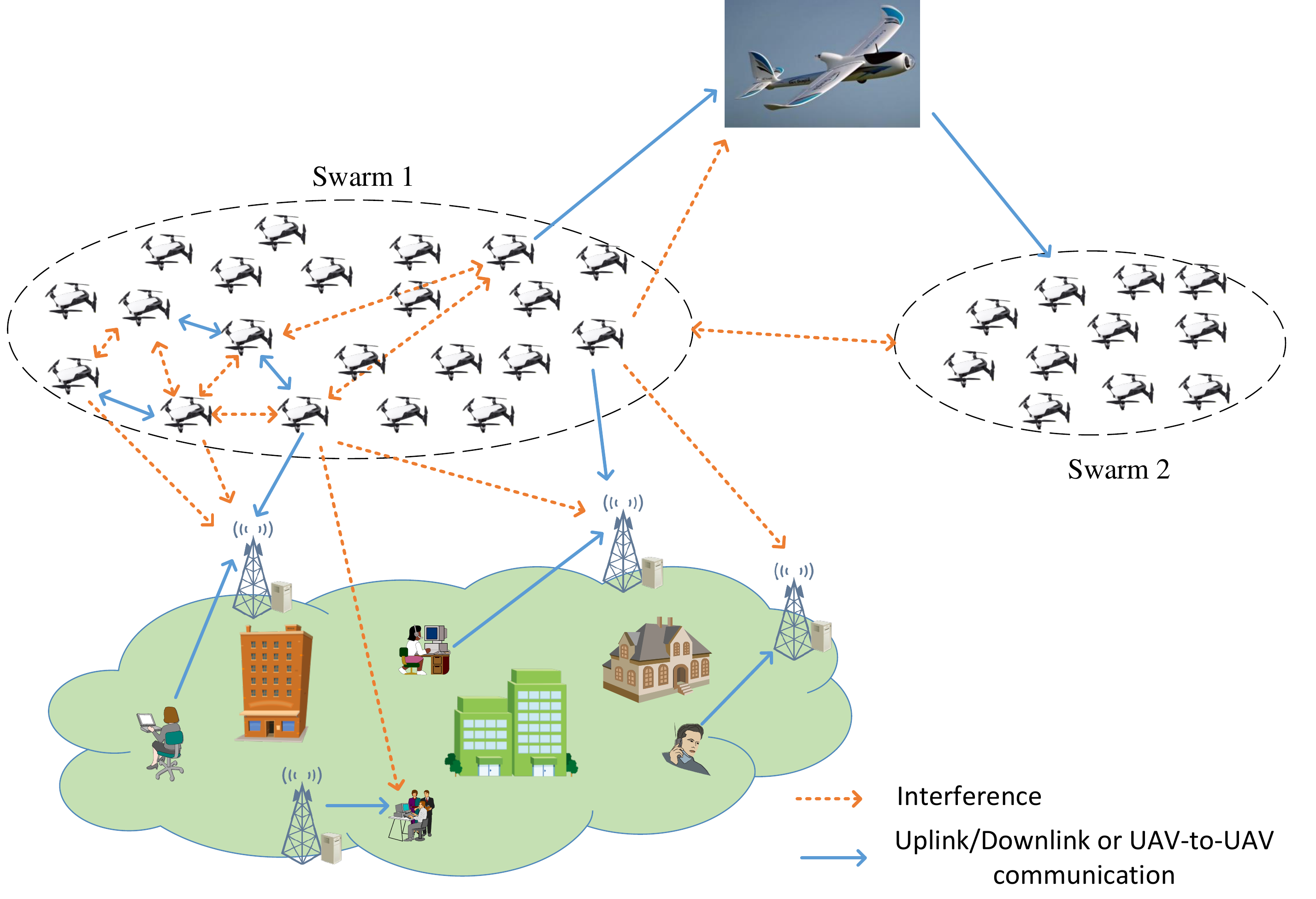}
\caption{The severe UAV-ground and UAV-UAV interferences.} \label{fig.3}
\end{figure}

One major challenge to ensure the low execution latency and high energy efficiency for the UAV swarm-enabled edge computing lies in the severe UAV-ground and UAV-UAV interference, as illustrated in Fig. \ref{fig.3}. Compared to conventional aerial edge computing systems, the interference in UAV swarm-enabled edge computing systems is greatly aggravated by the densely deployed UAVs in the swarm.

Lately MIMO becomes a very promising technique  to mitigate the strong intra-group interference and aerial-ground interference. Nevertheless, the existing pilot contamination issue becomes even worse with higher pilot requirements for massive MIMOs. How to deal with this issue is also challenging, since the channel characteristics of UAVs in the same swarm are rather similar and may not show much difference.


\subsection{Robust Automatic Networking for Large-Scale UAV swarm}

Large-scale UAV automatic networking is of vital importance in UAV swarm-enabled edge computing networks. At the beginning, a large number of UAVs in the surrounding area can form a swarm through either manual remote control operations or autonomously embedded commands. During the mission period, in order to maintain the stability of the FANET, each UAV can use the sensor data (e.g., the data collected from task zone) and information (e.g., other UAVs' velocity and heading angle) received through intra-swarm wireless network to coordinate its movements. However, in practice, the severe intra-swarm wireless interference, the uncertainty of wireless channel and data processing latency will inevitably cause response delay of UAV, which impairs the stability of the UAV swarm. Therefore, more research efforts are needed to investigate the  effective robust automatic networking technologies for keeping stability of large-scale UAV swarm.

\subsection{Balanced Computation Offloading}

In some application scenarios like battle fields, earth and environment monitoring, and disaster management and good delivery of UAV swarm-enabled edge computing, both the edge of the ground core network and remote cloud center are too far away to establish reliable connection. Hence, the ever commonly employed unbalanced computation offloading that offloads data flows to nearby powerful computation facilities without energy and computation ability constraints will no longer be applicable. It is important yet challenging to design efficient resource allocation strategies under balanced computation offloading mode that takes into account the trade-off between energy consumption and calculation speed, since the environment experienced by each UAV in a swarm is highly dynamic.

\subsection{Computing Security}

The constantly changed topology of FANET brings security risk to the collaborative computing among UAVs. Moreover, in most ``real-world'' networks,  GMESs or aerial computing platforms are either privately owned or controlled by different service providers. Computing task sharing among them carries  a great risk of privacy leakage. Therefore, computing security in UAV swarm-enable edge computing networks is of vital importance. To the best of our knowledge, there have been no studies on the data sharing security issues of UAV swarm-enable edge computing networks. The physical layer security techniques and the distributed federated learning provide a promising pursuit in this interesting direction.

\subsection{Online Path Planning}

While path planning has been extensively investigated for traditional UAV-enabled MEC networks, such an issue in UAV swarm-enabled edge computing networks faces new challenges. On one hand, the correlation and collaboration among UAVs require that each UAV plans its path based on global observations  rather than the local one. On the other hand, online path planning becomes inevitable to ensure the powerful task execution capability of UAV swarm, since the unexpected computational task arrival may pose great uncertainty to offline path planning.



\subsection{Energy-Efficient Multi-UAV Coordination}

As mentioned above, in the execution of the practical task, multi-UAV coordination is an extremely complex and challenging issue. When performing the cooperative mission planning, it inevitably encounters an NP hard problem. As the problem size (the number of UAVs, the targets, etc.) increases,  the solution space of the problem expands exponentially and explosively, which requires a large amount of computation and energy. Moreover, the communication network of UAV swarm is impaired by the intricate and changing environment. Hence, for the energy constrained UAVs, how to obtain energy-efficient multi-UAV coordination is a challenging and open issue.

\section{Conclusions}

In this article, we presented an overview of edge computing enabled by UAV swarm. The appealing advantages, potential application scenarios, and the network architectures were  presented  for  UAV swarm-enabled edge computing. The key implementation considerations were provided in order to facilitate the  understanding on practical application of UAV swarm edge computing. Moreover, the promising technologies to enable future edge computing systems with UAV swarm were discussed. Key challenges and open issues were outlined to provide an enlightening guidance for future research directions. It is concluded that the research on UAV swarm-enabled edge computing is in its infancy and extensive research efforts are needed to bring it to maturity.

\end{document}